\documentclass[pdftex,12pt]{article}
\usepackage{jheppub,amsmath,amsfonts,amsbsy,latexsym,jheppub,amssymb,amscd,amstext}
\usepackage{pst-node}
\usepackage{dsfont}
\usepackage{overpic,youngtab}
\pdfoutput=1

\def\be{\begin{equation}}
\def\ee{\end{equation}}
\def\bea#1\eea{\begin{align}#1\end{align}}
\def\pd{\partial}
\def\a{\alpha}
\def\b{\beta}
\def\g{\gamma}
\def\d{\delta}
\def\m{\mu}
\def\n{\nu}

\def\l{\lambda}

\def\s{\sigma}
\def\e{\epsilon}

\def\bma{\begin{pmatrix}}
\def\ema{\end{pmatrix}}

\def\bi{\begin{itemize}}
\def\ei{\end{itemize}}

\title{\boldmath Quasi-local energy and compactification}

\preprint{IFT-UAM/CSIC-18-041 \ FTUAM-18-9}

\author[a]{Enrique Alvarez,}
\author[a]{Jesus Anero,}
\author[a]{Guillermo Milans del Bosch} 
\author[a]{and Raquel Santos-Garcia}

\affiliation[a]{Departamento de F\'{\i}sica Te\'orica and Instituto de F\'{\i}sica Te\'orica, IFT-UAM/CSIC\\Universidad Aut\'onoma, 20849 Madrid, Spain}

\emailAdd{enrique.alvarez@uam.es} 
\emailAdd{jesusanero@gmail.com}
\emailAdd{guillermo.milans@csic.es}
\emailAdd{raquel.santosg@uam.es}

\abstract{Based on the quasi-local energy definition of Brown and York, we compute the integral of the trace of the extrinsic curvature over a codimension-2 hypersurface. In particular, we study the difference between the uncompactified Minkowski spacetime and the toroidal Kaluza-Klein compactification. For the latter, we find that this quantity interpolates between zero and the value for the uncompactified spacetime, as the size of the compact dimension increases. Thus, the quasi-local energy is able to discriminate between the two spacetimes. 
}
\begin{document}

\maketitle
\flushbottom
\thispagestyle{empty}
\newpage
\setcounter{page}{1}

	\section{Introduction}
One of the first problems encountered when trying to understand quantum effects in gravity (confer \cite{Alvarez}  and references therein) is that there is no available energetic argument in order to determine the ground state of the theory. This is one of the many aspects in which gravity differs from the other fundamental interactions, where there is a well-defined hamiltonian which is supposed to be minimized by the vacuum of the theory.
\par
In the case of gravitation, any Ricci-flat spacetime is a priori a valid candidate to a ground state and it is believed that different asymptotics are in different energy sectors. Therefore, it does not have physical sense to compare the respective energies, even in the few cases in which they can be computed (essentially the ADM or the Bondi mass) \cite{ADM,Bondi}. 
\par
One would like to have a criteria to discriminate, for example, between a $n$-dimensional Ricci-flat spacetime from another $n$-dimensional Ricci-flat spacetime with some dimensions compactified; that is a Kaluza-Klein \cite{KK,Klein} type of vacuum. 
E. Witten \cite{Witten} has been able to show that the five-dimensional Kaluza-Klein vacuum is semiclassically unstable; but no general energetic argument is available.
\par

Recently, however, a more general concept of gravitational energy has been proposed (see \cite{Szabados} for a recent review), namely {\em quasi-local energy} (QLE). The fact that it is a quasi-local quantity makes it suitable to compare spacetimes with different asymptotics. There are several definitions of QLE in the literature \cite{Bartnik,Hawking,Hayward,Penrose,Liu,Wang,Brown:1992br}. Here we follow the one by Brown and York \cite{Brown:1992br}. The main idea is to associate to a given hypersurface of a spacetime, $\Sigma \hookrightarrow \mathcal{M}$, the integral of the trace of the second fundamental form. Schematically\footnote{We are working in units where $\kappa^2=1$.},
\be
Q(\Sigma)\equiv \int_\Sigma K-E_0,
\ee
where the zero-point energy $E_0$ is computed by an isometric embedding of the hypersurface in $\mathbb{R}^3$ (or else in $M_4$ in other versions \cite{Wang}).
\par
 The aim of the present paper is to begin the exploration of the QLE in toroidal spacetimes and compare it to the corresponding uncompactified spacetime. In this preliminary investigation we are going to discuss very simple examples, for which we believe the discussion of the zero-point energy to be less relevant.
\par
To set up our notation, consider a codimension-$p$ hypersurface $\Sigma$ embedded in an ambient spacetime $\mathcal{M}$ of dimension $n$ and Lorentzian signature, $ \Phi: \Sigma \rightarrow \mathcal{M}$. Let $ y^\a $ and $ x^a $ be two coordinate systems on $ \mathcal{M} $ and $ \Sigma $, respectively, with $ \a=1,...,n $ and $ a=1,...,m$, where $ m=n-p $ is the dimension of $ \Sigma $. The embedding is defined by the equations
\be
\Phi : y^\a= y^\a(x^a).
\ee 
 Denoting by $g_{\m\n}(y)$ the metric in the ambient manifold, the induced metric on the hypersurface is given by
\be
h_{ab}(x)\equiv g_{\a\b}\left(y\left(x\right)\right){\pd y^\a\over \pd x^a}{\pd y^\b\over\pd x^b}.
\ee
The $m$ vectors on the tangent space to the ambient manifold, $\mathcal{T}(\mathcal{M})$, tangent to the hypersurface are given by
\be
t_a^\a\equiv {\pd y^\a\over \pd x^a},
\ee
\par
In the useful reference \cite{Eisenhart} it is proved the fact that if $h\equiv \text{det} (h_{ab})\neq 0$, then there are $p$ (as many as the codimension of the hypersurface) real mutually orthogonal vectors normal to $\Sigma$, none of which are null. Let us denote them by $n_A\in \mathcal{T}(\mathcal{M})$, $A=1,..., p$.
\be
n_A\cdot n_B=\e(A) \ \d_{AB} 
\ee
where $\e(A)=\pm 1$\footnote{Our metric conventions are $(+---...)$}. The generalization of the second fundamental form is the set of $p$ symmetric tensors given by
\be
K^A_{ab}\equiv n^A_\a t_b^\l \nabla_\l t^\a_a=-t_b^\l t^\a_a \nabla_\l n^A_\a,
\ee
where the orthogonality $n^A\cdot t_a=0$ has been used. 
\par 
We are interested in the integral
\begin{equation}\label{key}
\int_\Sigma \ \sqrt{h} \ K^\a n_\a dS,
\end{equation}
where $ dS $ is the surface element of the hypersurface $ \Sigma $ and 
\begin{equation}\label{QLE}
K^\a=h^{ab}K^\a_{ab}\equiv h^{ab}K^A_{ab} n^\a_A.
\end{equation} 
\par
The paper is organized as follows. In the next section we compute \eqref{QLE} in the case of 5-dimensional flat spacetime. In section 3 we repeat the calculation for the compactified spacetime $ M_4 \times S_1 $. Then, in section 4 we study the stationary points of the QLE before we end up with some conclusions. 

\section{Codimension-2 spheres in $M_5$}
Consider a codimension-2 spacelike hypersurface in 5-dimensional flat space 
\begin{equation}\label{key}
ds^2=\eta_{\a\b}dx^\a dx^\b.
\end{equation}
Let the hypersurface be a 3-sphere defined by the embedding
\bea
&y_1=T,\nonumber\\
&\sum _{i=2}^{i=5} (y_i)^2\equiv  L^2,
\eea
where latin indices $ i,j,... $ denote spatial coordinates. The normal vectors are given by
\be
n_A\equiv \left({\pd\over \pd t},{y^i\over L}{\pd\over \pd y^i}\right).
\ee
\par
In this setup, it is plain that the only normal vector with non-vanishing derivative is the last one
\be
n\equiv n_2= {y^i\over L}{ \pd\over \pd y^i}.
\ee
It yields
\be
\nabla_\b n^\a={L^2 \d^\a_\b- y^\a y_\b\over L^3}.
\ee
We have to project this on the tangent space using the tangent vectors $ t^\a_a $. In spherical coordinates, the hypersurface can be parametrized as follows
\bea
&y_2= L\sin\,\theta_1 \, \sin\,\theta_2 \,\sin\,\theta_3\nonumber\\
&y_3= L\sin\,\theta_1 \, \sin\,\theta_2 \,\cos\,\theta_3 \nonumber\\
&y_4= L\sin\,\theta_1 \cos\,\theta_2 \nonumber\\
&y_5= L\cos\,\theta_1  
\eea
so that the induced metric reads
\be
d\s^2 = -L^2\left(d\theta_1^2+\sin^2\,\theta_1 \, d \theta_2^2 + \sin^2\,\theta_1 \,\sin^2\,\theta_2 \,d\theta_3^2\right)
\ee
It follows that the tangent vectors $t^\a_a$ take the form
\bea
&t_{\theta_1}=L\left(0,\cos\,\theta_1\,\sin\,\theta_2 \,\sin\,\theta_3,\cos\,\theta_1\, \sin\, \theta_2\,\cos \theta_3,\cos\,\theta_1 \, \cos \theta_2,-\sin \theta_1\right),\nonumber\\
&t_{\theta_2}=L\left(0,\sin\,\theta_1\,\cos\,\theta_2 \,\sin\,\theta_3,\sin\,\theta_1\, \cos\, \theta_2\,\cos \theta_3, -\sin \theta_1 \, \sin \theta_2,0\right),\nonumber\\
&t_{\theta_3}=L\left(0,\sin\,\theta_1\,\sin\,\theta_2 \,\cos\,\theta_3,-\sin\,\theta_1\, \sin\, \theta_2\,\sin \theta_3, 0,0\right),
\eea
and satisfy
\be
y_\a t^\a_a=0. \label{orthogonal}
\ee
Their norm $t^2=\eta_{\a\b}t^\a t^\b$ is given by
\be
t_{\theta_1}^2=-L^2 \, ; \quad t_{\theta_2}^2=-L^2 \, \sin^2 \theta_1\, ;  \quad t_{\theta_3}^2=-L^2 \, \sin^2 \theta_1 \, \sin^2 \theta_2\, .
\ee
Hence,\footnote{The Gauss-Codazzi equations, which relate the ambient Riemann tensor $R_{\a\b\g\d}$ projected on the hypersurface with the Riemann tensor corresponding to the induced metric,$R_{abcd}[h]$
\be
t_a^\a t_b^\b t_c^\g t_d^\d \,R_{\a\b\g\d}=R_{abcd}[h]-\sum_{A=1}^{A=p}\e(A)\,\left(K_{ac}^A K_{bd}^A- K_{ad}^A K_{bc}^A\right).
\ee
provide a useful check of our computations.
}
\be
K^A_{ab}=\left( 0, - {1\over L}\,\d_{\a\b}t^\a_a t^\b_b  \right)
\ee
The integrand of \eqref{QLE} is then given by
\be
K^\a n_\a= h^{ab} \ K^\a_{ab}\ n_\a= \dfrac 3 L .
\ee
The integration measure in this coordinates takes the form 
\be
\sqrt{h} \ dS =  \ L^3 \sin^2 \,\theta_1 \, \sin \,\theta_2 \, d\theta_1  d\theta_2  d\theta_3 ,
\ee
so that the integral finally yields
\be
Q_{M_5}=\int_\Sigma \ \sqrt{h} \ K^\a n_\a  dS = 6 \pi^2 L^2.
\ee

\section{Codimension-2 spheres in  $M_4\times S_1$}
The metric of the ambient space is now
\be
ds^2=\eta_{\m\n} dy^\m dy^\n.
\ee
where the last coordinate is compact and has periodicity
\be
y_5=y_5+2\pi l,
\ee
and $l$ is the radius of the compact dimension. We assume the same algebraic surface as in the previous case, that is
\bea
y_1&=T \nonumber \\
\sum_{i=2}^{i=5} y_i^2 &= L^2.
\label{hyp2}
\eea
\par
In this case, the spacelike normal vector depends on the compact coordinate, but its expression is the same as in the previous case
\be
n={y_i\over L}{\pd\over \pd y^i}.
\ee
Also, the covariant derivative of the normal vector can still be written as
\be
\nabla_\b n^\a={L^2 \d^\a_\b- y^\a y_\b\over L^3}.
\ee
For this computation, we parametrize the hypersurface in cartesian coordinates
\bea
&y_2= x ,\nonumber\\
&y_3= y,\nonumber\\
&y_4= z ,\nonumber\\
&y_5= \sqrt{L^2 -x^2 -y^2 -z^2} .
\eea
The induced metric then reads
\be
h_{ab}=\dfrac{1}{L^2 -x^2 -y^2 -z^2} \left(
\begin{array}{ccc}
L^2-y^2-z^2& x y & x z \\
x y &L^2-x^2-z^2 & y z \\
	x z& y z &L^2-x^2-y^2\\
\end{array}
\right)
\ee
The tangent vectors still obey
\be
y_\a t_a^\a=0.
\ee
Explicitly, they read
\bea
&t_x =(0,1,0,0,\dfrac{-x}{ \sqrt{L^2 -x^2 -y^2 -z^2} }),\nonumber\\
&t_y =(0,0,1,0,\dfrac{-y}{ \sqrt{L^2 -x^2 -y^2 -z^2} }),\nonumber\\
&t_z =(0,0,0,1,\dfrac{-z}{ \sqrt{L^2 -x^2 -y^2 -z^2} }).\nonumber\\
\eea
The second fundamental form $K_{ab}^A$ then takes the form
\be
K_{ab}^1=0 \ ; \quad \quad  K_{ab}^2=\dfrac 1 L \left(
	\begin{array}{ccc}
		\frac{x^2}{L^2-x^2-y^2-z^2}+1 & \frac{x y}{L^2-x^2-y^2-z^2} & \frac{x z}{L^2-x^2-y^2-z^2} \\
		\frac{x y}{L^2-x^2-y^2-z^2} & \frac{y^2}{L^2-x^2-y^2-z^2}+1 & \frac{y z}{L^2-x^2-y^2-z^2} \\
		\frac{x z}{L^2-x^2-y^2-z^2} & \frac{y z}{L^2-x^2-y^2-z^2} & \frac{z^2}{L^2-x^2-y^2-z^2}+1 \\
	\end{array}
	\right)\ .
\ee
The integrand of \eqref{QLE} is again given by
\be
K^\a n_\a= h^{ab} \ K^\a_{ab}\ n_\a= \dfrac 3 L .
\ee
\par
\begin{figure}[h!!t]
	\begin{center}
		\includegraphics[width=0.7\textwidth]{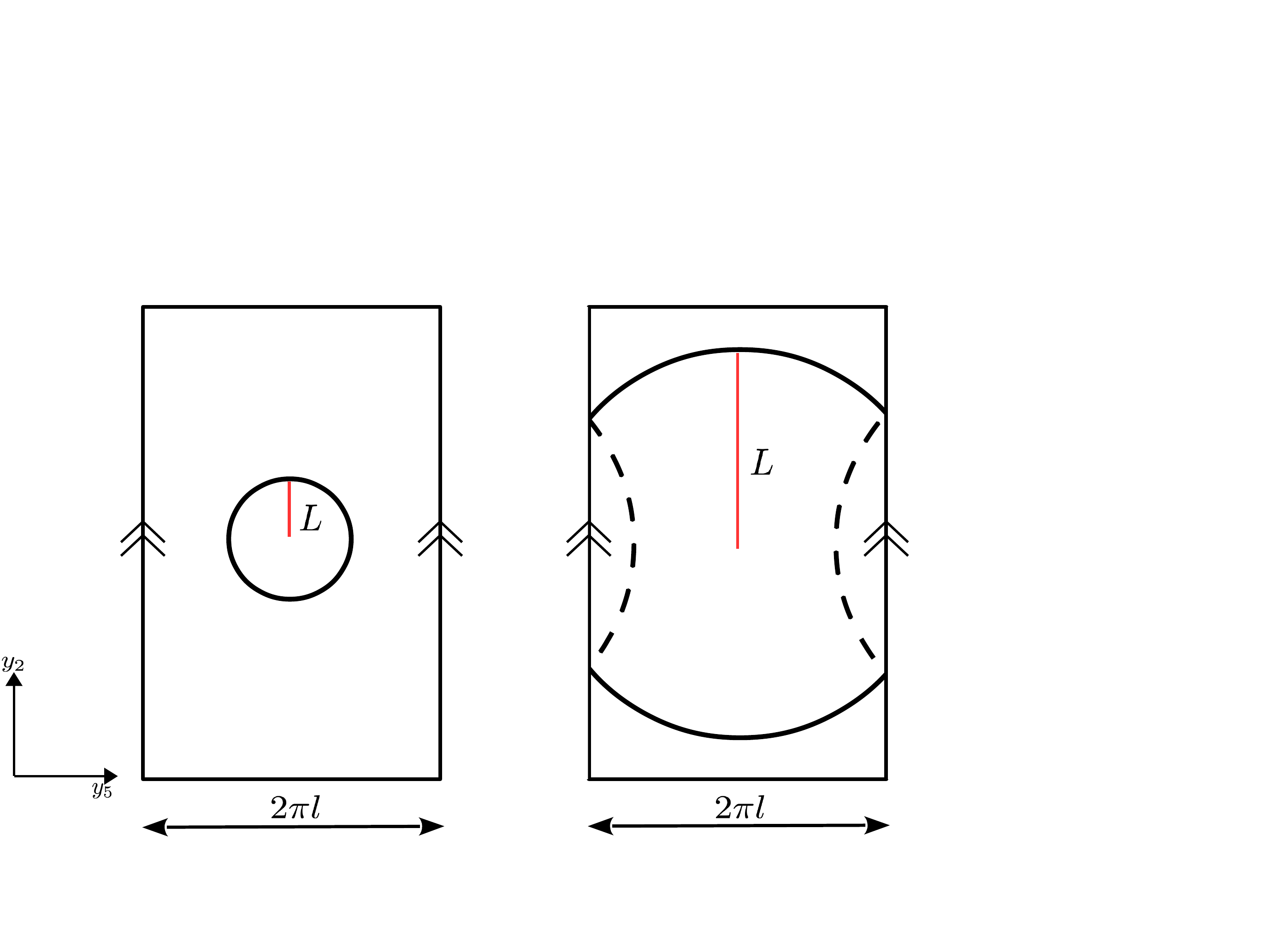}
	\end{center}
	\caption{Different types of hypersurface depending on whether  $L>\pi l$ or $L<\pi l$. For simplicity, we only show the compact dimension, $y_5$, and one extended dimension, $y_2$.}
	\label{fig3}
\end{figure}
One has to be careful with the integration range over the compactified coordinate. For small 3-spheres that completly lie within the compact dimension, that is with $L< l \pi$, the integration is done over the full hypersurface,  so that $-L \leq y_5 \leq L$. On the other hand, when $L> l \pi$, there are self intersections of the hypersurface, due to the periodicity of the compact dimension. Thus, the integration range is restricted to $- l \pi \leq y_5 \leq l \pi$, as can be seen in figure \ref{fig3}. We obtain
\bea
Q_{M_4 \times S_1}&= 6 \pi^2 L^2  \quad  \quad \text{for} \quad L \leq l \pi ,  \nonumber\\
Q_{M_4 \times S_1}&=    12 \pi^2 l \sqrt{L^2-\pi ^2 l^2}+ 12 \pi L^2 \tan ^{-1}\left(\frac{\pi  l}{\sqrt{L^2-\pi ^2 l^2}}\right) \quad \quad \text{for} \quad L > l \pi . 
\eea


\begin{figure}[h!!]
	\begin{center}
		\includegraphics[width=0.7\textwidth]{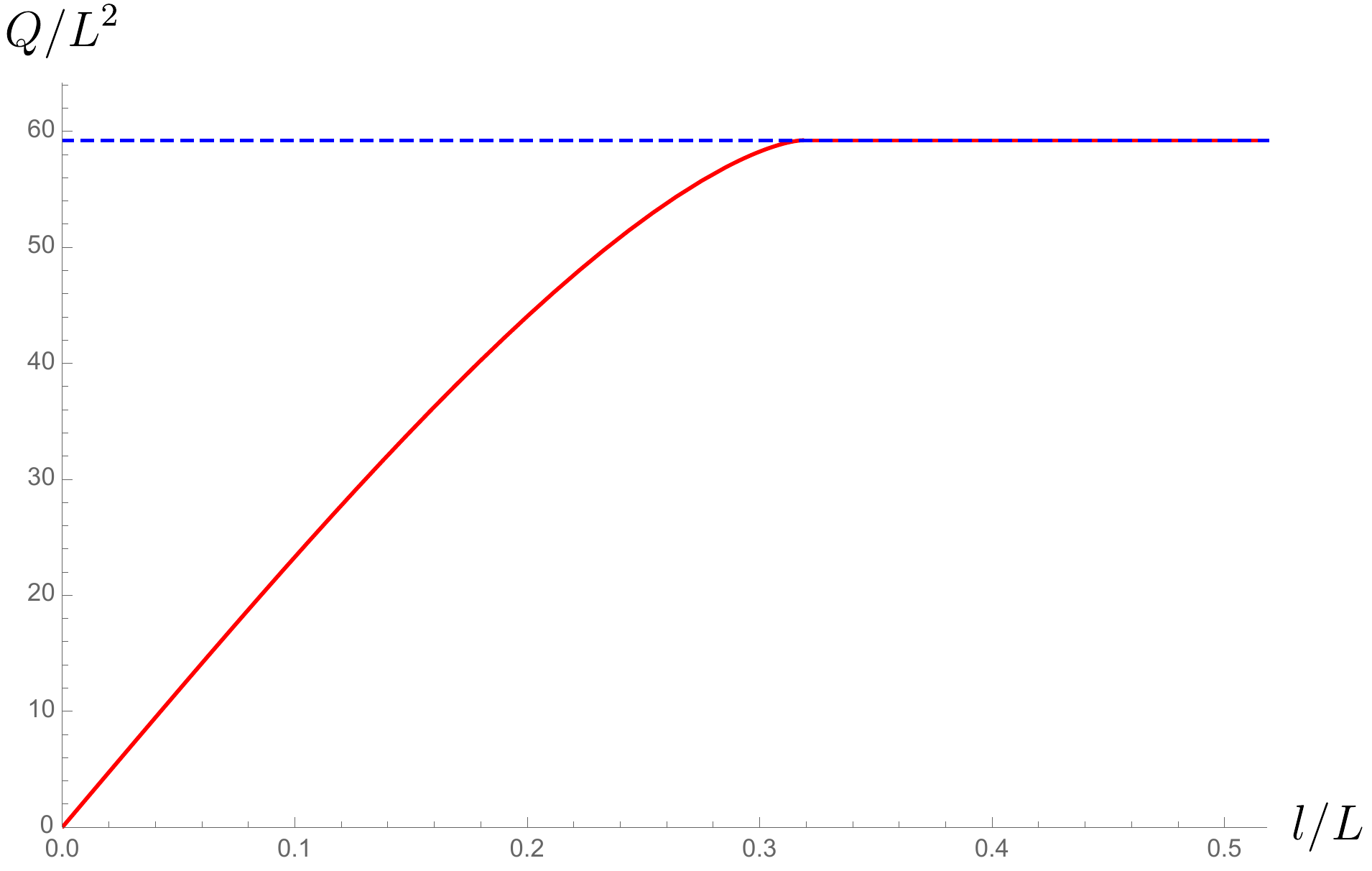}
	\end{center}
	\caption{Comparison between the QLE for $M_5$ (dashed blue) and $M_4 \times S_1$ (red).}
	\label{fig1}
\end{figure}

As expected, in the decompactification limit where $l \rightarrow \infty$ for any finite $L$, the QLE for $M_4 \times S_1$ is that of $M_5$. In fact, this happens whenever $L \leq l \pi $, since the hypersurface does not see the periodicity of the compact dimension; thus, the QLE cannot distinguish between $M_4 \times S_1$ and $M_5$. When $L> l \pi$, as can be seen in figure \eqref{fig1}, the QLE monotonically decreases to zero as $l \rightarrow 0$.
\section{Stationary points of the QLE}
Let us study the stationary points of the QLE integral under variations of the spacetime ambient metric, keeping fixed the equation for the embedding
\be
\d Q\equiv \d \int \sqrt{h}\,d^{n-2} x\, h^{ab} \,t_a^\a \,t_b^\b\, \nabla_\a n_\b=0.
\ee
From the normalization of the normal vectors we have
\be
g_{\a\b} n^\a_A n^\b_B=\eta_{AB} \Longrightarrow \d n^\a=-g^{\a\g} \d g_{\g\b} n^\b,
\ee
where from now onwards we will omit the label $A$ in the normal vectors. Orthogonality between normal and tangent vectors implies
\be
\d\left( n^\a g_{\a\b} t_a^\b\right)=0,
\ee
and note that
\be
\d n_\a = \d \left( g_{\a\b} n^\b\right)=0.
\ee
Let us define the  auxiliary tensor
\be
G^{\a\b}\equiv h^{ab} t_a^\a t_b^\b,
\ee
(remember that the tangent vectors we are using are not normalized), in such a way that
\be
G^{\a\b} g_{\a\b}\equiv G^\a_\a= h^{ab} h_{ab}=n-2.
\ee
\par
The determinant of the induced metric also varies
\be
\d h= h h^{ab} t_a^\a t_b^\b \d g_{\a\b}=h G^{\a\b} \d g_{\a\b},
\ee
because
\be
\d h_{ab}= t_a^\a t_b^\b \d g_{\a\b}\ ; \hspace{1cm}\d h^{ab}= - h^{ac} h^{bd} \d h_{cd}.
\ee
Thus, the variation of the QLE reads
\bea
&\d Q=\int \sqrt{h} d^n x\,\bigg\{{1\over 2}\d g_{\a\b} G^{ \a\b} h^{ab} t_a^\m t_b^\n \nabla_\m n_\n- h^{ac} h^{bd} t_c^\m t_d^\n \d g_{ \m\n} t_a^\a t_b^\b\nabla_\a n_\b+\nonumber\\
&-{1\over 2} G^{ \a\b} g^{\g\d} n_\g \left( -\nabla_\d \d g_{ \a\b}+\nabla_\a \d g_{ \b\d}+\nabla_\b \d g_{\a\d}\right)\bigg\}.
\eea
It is not possible in general to integrate by parts, because
\be
g\neq h.
\ee
It would be interesting to study classes of solutions to those integral-differential equations. In the particular case where the variation of the metric is assumed to be covariantly constant
\be
\nabla_\g \d g_{\a\b}=0,
\ee
the equations reduce to the much simpler condition
\be
K_{ab} t^{a \a} t^{b \b}={1\over 2} K G^{\a \b},
\label{KaKa}
\ee
where $K=K^\a n_\a$. For {\em umbilic surfaces} where the extrinsic curvature is proportional to the induced metric
\be
K_{ab}=\l h_{ab},
\ee
 \eqref{KaKa} reduces to
\bea
 \l h_{ab}t^{a\a} t^{b\b} &= {1\over 2}\l h^{ab} h_{ab} h_{cd}t^{c\a} t^{d\b}, \nonumber \\
 \l &= {\l \over 2} (n-2),
\eea
so that it implies 
\be
\l=0 \quad \text{or} \quad n=4.
\ee
We leave the study of more complex hypersurfaces for further investigation. 
\section{Conclusions}
We have begun to apply some preliminary ideas on quasi-local energy to the simplest instances of toroidal compactifications, and we have found somewhat surprisingly, that this observable can be sensitive to it. We take this fact as an encouragement to pursue this set of ideas, with the final objective in mind of being able to apply energetic arguments to the study of the ground state of fundamental physics including gravity.
 \par
In particular, when one dimension is allowed to compactify, we find that there is a runaway behavior of sorts, and the configuration that minimizes the QLE corresponds to this dimension disappearing completely. It is even possible that this behavior  is not unrelated to the old problem of stabilization of extra dimensions in a Kaluza-Klein setting (confer \cite{Brandenberger1, Branderberger2}, and references therein).
\par
We have also determined the equations that make stationary the QLE under arbitrary variations of the spacetime metric. They are quite complicated, but seem worthy of further consideration.
\par
We are aware that we are exploring uncharted waters here. Ours are only preliminary ideas. The r\^ole of the zero point energy, for example, has not been touched upon in our work.
\par
There are several lines of further work that can be pursued. It would be interesting to analyze the cases where not all the compact dimensions are contained in the hypersurface, as well as to study more general compact geometries. This includes, in particular,  the energetics of fluxes in non-trivial cycles \cite{Ibanez} as compared with the same geometry without the fluxes. Although we considered a simple example, it should be straightforward to generalize it to higher dimensional compact spaces.

	\section*{Acknowledgments}
One of us (EA) is grateful to the organizers of the workshop ``Mass in General Relativity", Piotr Chrusciel, Richard Schoen, Christina Sormani, Mu-Tao Wang, and Shing-Tung Yau for the kind invitation, as well as Luis \'Alvarez-Gaum\'e for the wonderful hospitality in the Simons Center, where the idea was conceived. GMB is supported by the project FPA2015-65480-P and RSG is supported by the Spanish FPU Grant No FPU16/01595. This work has received funding from the Spanish Research Agency (Agencia Estatal de Investigacion) through the grant IFT Centro de Excelencia Severo Ochoa SEV-2016-0597, and the European Unions Horizon 2020 research and innovation programme under the Marie Sklodowska-Curie grants agreement No 674896 and No 690575. We also have been partially supported by FPA2016-78645-P(Spain) and COST actions MP1405 (Quantum Structure of Spacetime). 


	\newpage
	\appendix
.	


\begin{thebibliography}{99}

\bibitem{Alvarez}
E.~Alvarez,
``Quantum Gravity: A Pedagogical Introduction To Some Recent Results,''
Rev.\ Mod.\ Phys.\  {\bf 61} (1989) 561.
doi:10.1103/RevModPhys.61.561


\bibitem{ADM} 
  R.~L.~Arnowitt, S.~Deser and C.~W.~Misner,
  ``The Dynamics of general relativity,''
  Gen.\ Rel.\ Grav.\  {\bf 40}, 1997 (2008)
  doi:10.1007/s10714-008-0661-1
  [gr-qc/0405109].
\bibitem{Bondi} 
  H.~Bondi, M.~G.~J.~van der Burg and A.~W.~K.~Metzner,
  ``Gravitational waves in general relativity. 7. Waves from axisymmetric isolated systems,''
  Proc.\ Roy.\ Soc.\ Lond.\ A {\bf 269}, 21 (1962).
  doi:10.1098/rspa.1962.0161



\bibitem{KK}
T.~Kaluza,
  ``Zum Unit�tsproblem der Physik,''
  Sitzungsber.\ Preuss.\ Akad.\ Wiss.\ Berlin (Math.\ Phys.\ ) {\bf 1921} (1921) 966
  [arXiv:1803.08616 [physics.hist-ph]].\\

\bibitem{Klein}
  O.~Klein,
  ``Quantum Theory and Five-Dimensional Theory of Relativity. (In German and English),''
  Z.\ Phys.\  {\bf 37}, 895 (1926)
  [Surveys High Energ.\ Phys.\  {\bf 5}, 241 (1986)].
  doi:10.1007/BF01397481\\

\bibitem{Witten}
E.~Witten,
``Instability of the Kaluza-Klein Vacuum,''
Nucl.\ Phys.\ B {\bf 195} (1982) 481.
doi:10.1016/0550-3213(82)90007-4

\bibitem{Szabados} 
L.~B.~Szabados,
``Quasi-Local Energy-Momentum and Angular Momentum in General Relativity,''
Living Rev.\ Rel.\  {\bf 12}, 4 (2009).
doi:10.12942/lrr-2009-4

\bibitem{Hawking} 
S.~Hawking,
``Gravitational radiation in an expanding universe,''
J.\ Math.\ Phys.\  {\bf 9}, 598 (1968).
doi:10.1063/1.1664615

\bibitem{Penrose}
R.~Penrose, 
``Quasi-Local Mass and Angular Momentum in General Relativity," 
Proceedings of the Royal Society of London. Series A, Mathematical and Physical Sciences, vol. 381, no. 1780, 1982, pp. 53-63. 

\bibitem{Bartnik} 
R.~Bartnik,
``New definition of quasilocal mass,''
Phys.\ Rev.\ Lett.\  {\bf 62}, 2346 (1989).
doi:10.1103/PhysRevLett.62.2346

\bibitem{Brown:1992br} 
J.~D.~Brown and J.~W.~York, Jr.,
Phys.\ Rev.\ D {\bf 47}, 1407 (1993)
doi:10.1103/PhysRevD.47.1407
[gr-qc/9209012].

\bibitem{Hayward} 
S.~A.~Hayward,
``Quasilocal gravitational energy,''
Phys.\ Rev.\ D {\bf 49}, 831 (1994)
doi:10.1103/PhysRevD.49.831
[gr-qc/9303030].

\bibitem{Liu} 
C.~C.~M.~Liu and S.~T.~Yau,
``Positivity of Quasilocal Mass,''
Phys.\ Rev.\ Lett.\  {\bf 90}, 231102 (2003)
doi:10.1103/PhysRevLett.90.231102
[gr-qc/0303019].


\bibitem{Wang} 
M.~T.~Wang and S.~T.~Yau,
``Quasilocal mass in general relativity,''
Phys.\ Rev.\ Lett.\  {\bf 102}, 021101 (2009)
doi:10.1103/PhysRevLett.102.021101
[arXiv:0804.1174 [gr-qc]].

\bibitem{Eisenhart}
L.P.Eisenhart, "Riemannian Geometry"
Princeton University Press (1964)

\bibitem{Brandenberger1}
R.~H.~Brandenberger and C.~Vafa,
``Superstrings in the Early Universe,''
Nucl.\ Phys.\ B {\bf 316} (1989) 391.
doi:10.1016/0550-3213(89)90037-0\\
\bibitem{Branderberger2}
S.~Watson and R.~Brandenberger,
``Stabilization of extra dimensions at tree level,''
JCAP {\bf 0311} (2003) 008
doi:10.1088/1475-7516/2003/11/008

\bibitem{Ibanez}
L.~E.~Ibanez and A.~M.~Uranga,
``String theory and particle physics: An introduction to string phenomenology,''
Cambridge University Press (2012)



\end{thebibliography}
\end{document}